\documentclass[aip,jap,12pt]{revtex4-1}
\usepackage{amssymb}
\usepackage{graphicx}
\usepackage{bm}
\usepackage{color}
\usepackage{epsfig}
\usepackage{amsfonts}

\begin{document}

\title{Quantum control of a model qubit based on a multi-layered quantum dot}

\author{Alejandro Ferr\'on}
\email{aferron@conicet.gov.ar}
\affiliation{Instituto de Modelado e Innovaci\'on Tecnol\'ogica
(CONICET-UNNE), Avenida Libertad 5400, W3404AAS Corrientes, 
Argentina}

\author{Pablo Serra}
\email{serra@famaf.unc.edu.ar}
\affiliation{Facultad de Matem\'atica, Astronom\'{\i}a y F\'{\i}sica,
Universidad Nacional de C\'ordoba and IFEG-CONICET, Ciudad Universitaria,
X5016LAE C\'ordoba, Argentina}

\author{Omar Osenda}
\email{osenda@famaf.unc.edu.ar}
\affiliation{Facultad de Matem\'atica, Astronom\'{\i}a y F\'{\i}sica,
Universidad Nacional de C\'ordoba and IFEG-CONICET, Ciudad Universitaria,
X5016LAE C\'ordoba, Argentina}

\begin{abstract}
In this work we present a model qubit whose basis states are eigenstates of a
multi-layered quantum dot. We show that the proper design of the quantum dot
results in qubit states that have excellent dynamical properties when a
time-dependent driving is applied to it. In particular, it is shown that a
simple sinusoidal driving is sufficient to obtain good quality Rabi oscillations
between the qubit states. Moreover, the switching between states can be
performed with very low leakage, even under off-resonance conditions. In this
sense, the quantum control of the qubit is robust under some
perturbations and achieved with simple means. 

\end{abstract}
\date{\today}

\maketitle

\section{Introduction}

Since Loss and DiVincenzo proposed the utilization of quantum dots as the
physical implementation of the qubit \cite{Loss1998}, there has been a huge
amount of work devoted to tackle the numerous and  subtle difficulties involved
in the problem. There are some excellent reviews
\cite{Reimann2002,Kouwenhoven2001} and books \cite{Braunstein2001} that
summarize the progress
experimented by the field, but it is extremely difficult to keep up with
the new developments. 

As much as any other proposal to implement a qubit, the spin degree of freedom
of an electron trapped in a quantum dot (QD), the original proposal made by Loss
and DiVincenzo,  must face a number of challenges owed to the intrinsic physics
that governs its behavior as a qubit.
Not to mention the challenges offered by other
physical implementations that try to catch the attention of the community
\cite{Braunstein2001}. 
The
double quantum dot scheme \cite{DiVincenzo2005} was devised to
circumvent the unavoidable decoherence induced by the interaction between the
angular momentum of the electron with the nuclear spins of the atoms 
that form the QD
\cite{Petta2005}. For this scheme, the realization of multiple qubit quantum
gates has been shown \cite{Brunner2011}. Nevertheless, since the coupling
between single QD's seems a bit problematic and maybe even more involved in the
double QD scheme, there has been a number of proposal showing that it is
possible to implement quantum control and refocusing techniques in single
quantum dots to restore the role of the single QD as a bona fide qubit. This,
together with techniques designed to distinguish between spatial states of the
trapped electron, make interesting again the search of new one-electron
structures that can be controlled with the exquisite precision required for the
quantum information tasks. 

The control of quantum systems, at least when the decoherence mechanisms are
absent or ``turned-off'', is implemented using pulses of external fields or
manipulating  an adequate parameter. A complete knowledge of the spectrum
allows the use of  ``navigating methods'' that make possible going from
(almost) any initial state to the desired target state \cite{Murgida2007}.
Nevertheless, the most used approach to achieve the switching between the two
qubit basis states is the optimal control theory \cite{Krotov1996,Sklarz2002}.
The application of Krotov's algorithm \cite{Krotov1996} usually leads to 
speedups of the transition time. This method has successfully been applied to
one- \cite{Rasanen2007,Rasanen2008} and two-electron quantum dots
\cite{Saelen2008}, allowing fast charge transfer with larger fidelities than
the obtained with simpler sinusoidal pulses. As much promising as the optimized
pulse method seems, the introduction of a complex modulation of the control
field necessarily introduces a host of new error sources  that have not been
properly analyzed. In this sense, achieving a good control of the switching
between states of a quantum dot without resorting to a complicated pulse
sequence is worth of study, and is the aim of this work.

Recently, it has been shown that multi-layered quantum dots can be designed to
selectively modulate the spatial extent of the electronic density of its
eigenstates \cite{Ferron2012,Tas2012}. This feature, together with the dipole
selection
rule permits, as we will show, the design of a quantum dot with two states that
can be switched, 
robust and efficiently, using only sinusoidal pulses. These two states are
the basis states of our model qubit. Using high-precision ab-initio numerical
calculations and exact solutions, where available, we aim to study the
spectrum, eigenstates of the quantum dot and dynamical properties of the qubit.

The paper is organized as
follows, in Section~\ref{model} a realistic multi-layered quantum dot model is
presented and qualitatively analyzed. In Section~\ref{speceigen} the properties
of the eigenvalues and eigenstates of the model are obtained. The study of the
spatial extent of the eigenstates allows the identification of potential qubit
basis states. The time evolution of the quantum state when the system is driven
by an external sinusoidal radio frequency field (rf) is the subject of
Section~\ref{sindriven}.  It is shown that the sinusoidal driving is enough to
obtain an excellent switching between the qubit basis states with very low
probability leakage, making unnecessary the utilization of complicated envelope
functions as is customary in optimal control theory. The stability of the qubit
oscillations is also tested considering the effect of off-resonance driving. In
Section~\ref{othersys} we briefly present a different model potential that also
define an excellent qubit, whose properties are analyzed through the lines
drawn in Sections~\ref{model}, \ref{speceigen}, and \ref{sindriven}. Finally,
we discuss our results in Section~\ref{dissc}.

\section{The model}
\label{model}

\begin{figure}
\begin{center}
\includegraphics[scale=0.5]{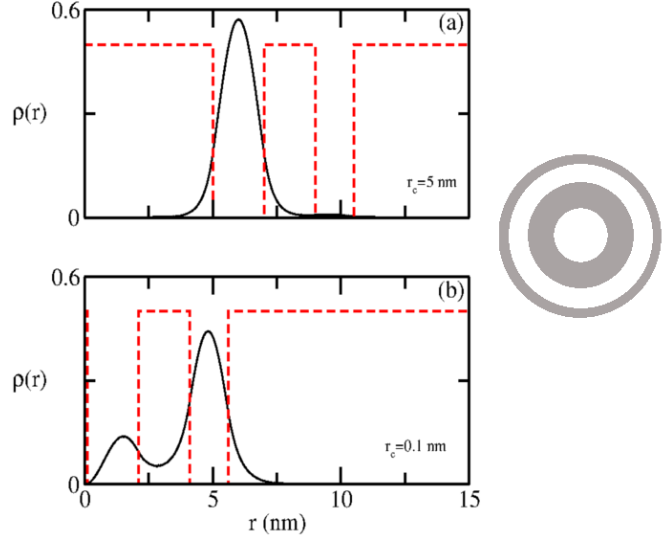}
\end{center}
\caption{\label{fig:one}(Color on-line) a) The electronic density for a
quantum state  well localized in the innermost potential well of a layered
quantum dot. 
The radial step-like potential
given by Equation~\ref{potrc}  is also shown (red dashed line). b) The
electronic density jumps from the innermost potential well to the outermost one
when the radius of the quantum dot central core is changed. The target-like
pattern to
the right of the figure corresponds to the cross-section of the quantum dot,
the grey zones correspond to both potential wells, while the central core, and
the barrier are depicted with white.
}
\end{figure}

In the Effective Mass Approximation, the Hamiltonian of trapped particles
assumes a simple form since the many-body interactions are reduced to a
bounding potential. In a spherical layered quantum dot, where each layer is
made of a different material, the bounding potential is given by the conduction
band off-sets of each material. Figure~\ref{fig:one} shows, schematically, the
radial profile of the bounding potential for a quantum dot made of two
different materials. The basic structure consists of a central core and two
wells separated by a barrier \cite{Ferron2012}. 

Despite its simplicity, the piecewise  potential takes into account a
number of experimental features and allows the formulation of accurate and
simple models for many nano-structures. The current semiconductor technology
permit the fabrication of layered structures where the radius of each layer can
be tuned with great precision. Figure~\ref{fig:one} also shows the qualitative
behavior of a given eigenstate, when the central core is wide enough the
electronic density is mostly located in the inner potential well, conversely,
when the radius of the central core is diminished the electronic density jumps
to the outermost potential well
for  a certain critical value. So, changing the quantum dot architecture is
equivalent to choose between different spectra and sets of eigenstates, whose
physical properties can be dramatically changed just altering the design of the
quantum dot. In particular, the change in the spatial extent of low lying
eigenstates and the modulation of the oscillator strength associated to these
eigenstates was analyzed in the work by Ferr\'on {\em et al.} \cite{Ferron2012}.

More precisely, in this work the bounding potential considered is given by

\begin{equation}
\label{potrc}
V_c(r) = \left\lbrace 
\begin{array}{lll}
V_0,& \quad & r < r_c; (ZnS), \\
0, & &  r_c\leq r<r_{1}; (CdSe),\\
V_0,& & r_1\leq r <r_2; (ZnS),\\
0,& & r_2\leq r <r_3; (CdSe),\\
V_0,& \quad & r \geq r_3 ; (ZnS).
\end{array}
\right.
\end{equation}
where $V_0=0.9$ $eV$ which corresponds
to the band offset between CdSe and ZnS, while the effective masses are
$m^{\star}_{e,CdSe} =0.13m_e$
and $m^{\star}_{e,ZnS} = 0.28m_e$, $m_e$ is the mass of the bare electron
\cite{Zhang2006}.

Since we are interested in the properties of a one electron quantum dot when it
is driven by an external field, the Hamiltonian takes the form

\begin{equation}\label{hamiltoniano1}
H=H_0+V_{ext}(\vec{r},t)
\end{equation}

\noindent where

\begin{equation}\label{hamiltoniano2}
H_0 = -\frac{\hbar^2}{2} \nabla \left(\frac{1}{m(\mathbf{r})}\right)\nabla +
V_c(r),
\end{equation}
where $m(\mathbf{r})$ is the position-dependent effective mass of the electron, 
and the bounding potential is given in \ref{potrc}. The kinetic energy term in
Equation~\ref{hamiltoniano2} preserves the Hermitian character of the
Hamiltonian  operator when the mass is position dependent\cite{Stern1984}. In
the present case the mass a step-like function of the radial coordinate.

\section{Spectrum and eigenstates}
\label{speceigen}

The spectrum and eigenstates of Hamiltonian~\ref{hamiltoniano2} can be obtained
exactly or numerically. In any case, the problem has spherical symmetry so
the eigenvalues depend on two quantum numbers, $(n_r,\ell)$, where $n_r$ is
radial quantum number, or the number of nodes of the radial eigenfunction, and
$\ell$ is the orbital angular momentum quantum number. On the other hand, note
that since the exact solution of the eigenvalue problem involves the roots of
transcendental equations, it can be difficult to figure out how many
quasi-degenerate eigenvalues has the problem. This difficulty is particularly
cumbersome for large values of $r_c$. Conversely, the variational methods
detect fairly well almost degenerate eigenvalues, so the data shown in this
Section was double-checked comparing the results from the numerical and
analytical procedures,

The numerical solutions were obtained using B-splines basis sets,
which are well suited to implement the boundary conditions imposed by the
step-like nature of the potential and the effective mass. Besides, B-splines
results are very accurate in comparison with calculations based on Gaussian,
Hylleraas and finite-element basis sets \cite{Serra2012}.

To use the B-splines basis, the  normalized one-electron
orbitals are given by
\begin{equation}\label{phi-bs}
\phi_{n}({r}) = C_n \, \frac{B^{(k)}_{n+1}(r)}{r}  \,;\;\;n=1,\ldots
\end{equation}

\noindent where $B^{(k)}_{n+1}(r)$ is a B-splines polynomial of order $k$.
The numerical results  are obtained by defining a cutoff radius $R$, and
then the interval $[0,R]$ is divided into $I$ equal subintervals.
 B-spline polynomials \cite{deboor} (for a review
of applications of B-splines polynomials in atomic and molecular physics,
see ref. \cite{bachau01})  are piecewise polynomials defined by a
sequence of  knots $t_1=0\leq t_2\leq\cdots \leq t_{2 k+I-1}=R$
and the recurrence relations

\begin{equation}\label{bs1}
B_{i}^{(1)}(r)\,=\,\left\{ \begin{array}{ll} 1 & \mbox{if}\,t_i\leq r <
t_{i+1}   \\
0 &\mbox{otherwise,}  \end{array}  \right. \,.
\end{equation}

\begin{equation}\label{bsrr}
B_{i}^{(k)}(r)\,=\,\frac{r-t_i}{t_{i+k-1}-t_i}\,B_{i}^{(k-1)}(r)\,+\,
\frac{t_{i+k}-r}{t_{i+k}-t_{i+1}}\,B_{i}^{(k-1)}(r)\; (k>1)\,.
\end{equation}

\noindent In this work, we use the standard  choice for the knots  in
atomic physics  \cite{bachau01} $t_1=\cdots=t_k=0$ and
$t_{k+I}=\cdots=t_{2k+I-1}=R$. We choose an equidistant distribution of
inside knots. The constant $C_n$ in Eq.(3) is a normalization
constant obtained from the condition $\langle n | n \rangle=1$,

\begin{equation}\label{nor-c}
C_n = \frac{1}{\left[ \int_0^{R_0} \, \left(B^{(k)}_{n+1}(r) \right)^2 \,dr
\right]^{1/2}} \,.
\end{equation}

Because $B_1(0)\ne0$ and $B_{I+k-1}(R)\ne0$, we have $N=I+k-3$ orbitals
corresponding to $B_2,\ldots,B_{I+k-2}$. In all the calculations we used
the value $k=5$, and,
 we do not write the index $k$ in the eigenvalues and coefficients.

\begin{figure}
\begin{center}
\includegraphics[scale=0.5]{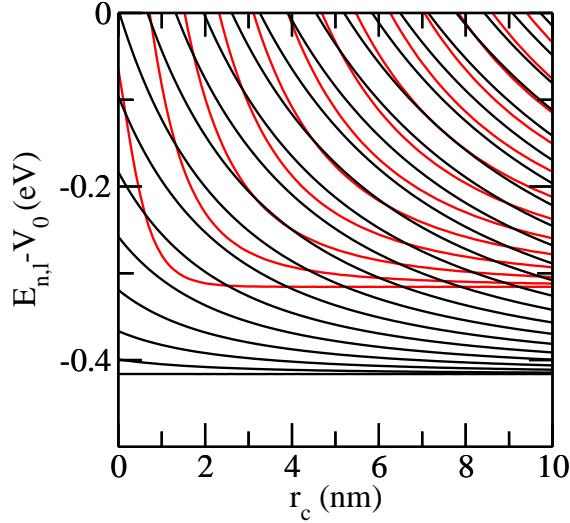}
\end{center}
\caption{\label{fig:two}(Color on-line) The spectrum of a typical layered
quantum dot as a function of the inner core radius, $r_c$. The radius of each
layer is, $r_1=r_c+0.8$, $r_2=r_1+3.5$, and $r_3=r_2+1$, all in nanometers.
The black solid lines correspond (from bottom to top) to  eigenvalues with
quantum numbers $(n_r=0,\ell), \; \ell=0,1,2,\ldots,22$, while the red solid
lines correspond to (from bottom to top) eigenvalues with quantum numbers
$(n_r=1,\ell), \; \ell=0,1,2,\ldots,12$. 
}
\end{figure}

Figure~\ref{fig:two} shows the spectrum of a double-well quantum dot as a
function of the central core radius, the width of the two wells and the barrier
are kept constant, so  $r_1= r_c + 0.8 \mbox{nm} $, $r_2 = r_1+ 3.5
\mbox{nm}$ and $r_3=r_2+1 \mbox{nm}$. As can be appreciated, the spectrum is
quite complicated and there is not a couple of energy values well separated
from the others, which is a common criterion to identify possible basis
eigenstates for a qubit. As we will show, the availability of a couple of
eigenstates that can be easily switched with a simple pulse depends not only on
the characteristics of the spectrum, but also in the eigenstates spatial extent.

\begin{figure}
\begin{center}
\includegraphics[scale=0.5]{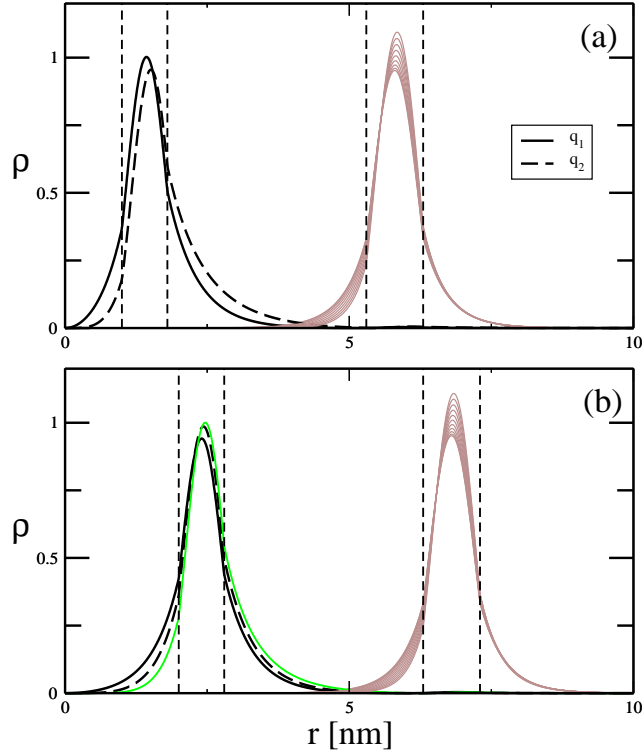}
\end{center}
\caption{\label{fig:three}(Color on-line) Electronic densities for the ground
and excited states for two different
CdSe/ZnS devices. (a) Device 1, 
(b) Device 2. The electronic densities of the eigenstates corresponding to the
qubit states are denoted with $q_1$ and $q_2$, which have quantum numbers
$(1,0)$ and $(1,1)$, respectively. 
. The black dashed vertical lines show the positions of
the barriers and
wells of the quantum dot. Both Devices are particular cases of the
layered quantum dot whose spectra are depicted in Figure~\ref{fig:two}.
}
\end{figure}

Figure~\ref{fig:three} shows the electronic density corresponding to all the
bounded eigenstates of two similar devices. Panel a) corresponds to a device
with only two eigenstates well localized in the innermost potential well
, while
panel b) shows a device with three eigenstates localized in the innermost
potential well. The two devices only differ in the central core radius size, it
is bigger for the b) case. The radii of both devices are, Device 1:$r_c=1$ nm,
$r_1=1.8$ nm,   $r_2=5.3$ nm and
$r_3=6.3$ nm. 
Device 2: $r_c=2$ nm, 
$r_1=2.8$ nm,  $r_2=6.3$ nm and $r_3=7.3$ nm.

In both cases the eigenstates  localized in the inner potential
well have different angular
momentum quantum numbers, in a) $\Delta \ell=1$. Actually,
in the Device 1 case, the
qubit lower state has quantum numbers $n_r=1$ and $\ell=0$, while the higher one
has $n_r=1$ and $\ell=1$, while in the Device 2 case the third state has $n_r=1$
and $\ell=2$. 
Figure~\ref{fig:three} shows two remarkable facts, i) the
number of eigenstates localized in a given potential well can be chosen for
realistic quantum dots parameters and, ii) the number of states of election
is pretty stable against small changes in the device's parameters (see also
Figure~\ref{fig:two}).  However, as
we will show latter a change in the number of states localized in the potential
well of interest can produce a huge change in the dynamical behavior once a
external driving is applied to the device. 

On the other hand, despite that we mostly present results for  particular sets
of parameters (effective masses, radii, etc.) the physical traits that, at
some extent, guarantee the existence of a small number of eigenstates well
localized in a multi-well potential are fairly general. Indeed, later on we
will show a example with a continuous bounding potential whose spectrum,
eigenstates and dynamical behavior are strikingly similar to those of the
devices with step-like potentials.

\section{Sinusoidal driving of the electron}
\label{sindriven}

The most simple non-trivial driving, both from an numerical and experimental
point of view, that can be applied to the trapped electron is  
\begin{equation}\label{eq:drive}
V_{ext}(\vec{r},t)=A_0\cos(\omega t)z
\end{equation}
where $A_0$ and  $\omega$  are the strength and the frequency of the driving,
respectively. The time-dependent potential, Equation~\ref{eq:drive}, models the
effect of a time-periodic spatially constant electric field applied in the $z$
direction. As the potential Equation~\ref{eq:drive} depends on $z$, the dipole
selection
rules impose transitions between eigenstates that differ in orbital angular
momentum, $\Delta\ell=\pm 1 $. 

The time evolution of the electron quantum state is governed by the
Schr\"odinger equation,
\begin{equation}\label{eq:schrodinger}
i \hbar \frac{\partial \Psi}{\partial t}=H\Psi
\end{equation}
where $\Psi$ is the quantum state and $H$ is the Hamiltonian in
Equation~\ref{hamiltoniano1}. 
Since we want to analyze the behavior of the devices presented in the previous
Sections as qubits, we will study the electron quantum state evolution, taking
as initial condition the lowest eigenstate that is localized in the innermost
potential well, from now on the $|0\rangle$ state of our putative qubit, this
state has $\ell =0$. The other qubit basis state, $|1\rangle $ is the
eigenstate with $\ell=1$ that is  also localized  in the innermost potential
well. Ideally, to qualify as a qubit, a physical system
 would perfectly switch between the two basis
states under the appropriate driving, {\em i.e.} if $|c_{q1}|^2$ and 
$|c_{q2}|^2$ are the time dependent probabilities that the electron is in the
$|0\rangle$ or in the $|1\rangle $ state, respectively, then
$|c_{q1}|^2 + c_{q2}|^2=1$. 

Except for ideal two level systems, there
is a finite probability that after a switching operation
$|c_{q1}|^2+c_{q2}|^2 < 1$, {\em i.e.} the qubit leaks probability. The
{\em leakage}, defined as $1 -(|c_{q1}|^2+c_{q2}|^2)$, is a good measure to
judge
the performance of a given system as a qubit \cite{Ferron2010}. In actual
physical systems, the
leakage has a twofold origin, the driving used to switch between the basis
states produces transitions to other levels besides the ones of interest, and
the interaction with the environment. In this work we will only analyze the
former without considering the possibility of ionization, so the electron
remains bounded while the driving is applied.

The material that follows is quite standard,
nevertheless we include it for the sake of completeness.
Writing  the quantum
state as a superposition of all the eigenstates 
\begin{equation}\label{eq:superposition}
\Psi(\vec{r},t)=\sum c_n(t)e^{-iE_nt/\hbar}\Phi_n(r)
\end{equation}
where
\begin{equation}
H_0\Phi_n(r)=E_n \Phi_n(r),
\end{equation}
is the eigenvalue problem, and replacing Equation~\ref{eq:superposition} in
the Schr\"odinger 
equation,~\ref{eq:schrodinger}, we get 
\begin{equation}
\sum c_n(t)e^{-iE_nt/\hbar}H\Phi_n=
i\sum \, \frac{\partial}{\partial t}\left( c_n(t)e^{-iE_nt/\hbar}\right)
\Phi_n .
\end{equation}
which can be solved for $c_k(t)$ using the orthogonality of the $\Phi_n$'s
\cite{Merzbacher}, 
\begin{equation}
i\frac{\partial c_k(t)}{\partial t}=
\sum_{n=0}^\infty c_n(t)\langle\Phi_k|V_{ext}|\Phi_n\rangle 
e^{i\omega_{kn}t}.
\end{equation}
Introducing the explicit for of the external driving, Equation~\ref{eq:drive},
we get that
\begin{equation}
\label{eq:et1}
i\frac{\partial c_k(t)}{\partial t}=A_0\cos(\omega t)
\sum_{n=0}^\infty c_n(t) Z_{kn}e^{i\omega_{kn}t}
\end{equation}
where $\omega_{kn}=(E_k-E_n)/\hbar$, and
\begin{equation}\label{eq:matrixz}
Z_{kn}=\langle\Phi_k|z|\Phi_n\rangle,
\end{equation}
clearly, the time-dependent probability that a given state, say $k$, is
occupied at time $t$ is given by $|c_k(t)|^2$

\begin{figure}[floatfix]
\begin{center}
\includegraphics[scale=0.5]{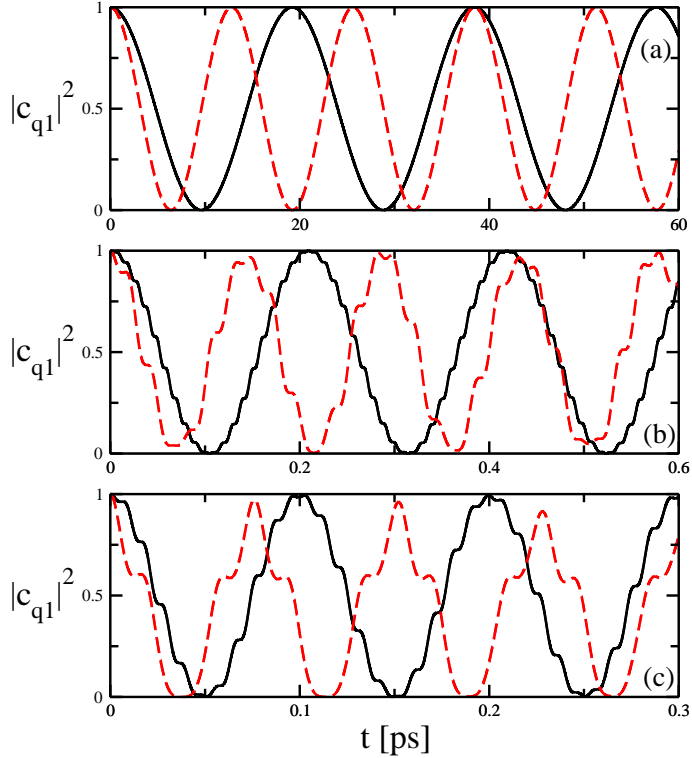}
\end{center}
\caption{\label{fig:four}(Color on-line) Time dependent probability of finding
the system in the lower state of the qubit ($|c_{q1}|^2$)
for different devices and  for an rf field pulse of strength $A_0$ at the
resonant frequency 
$\omega_{res}=(E_{q2} - E_{q1})/\hbar$ .
The system is prepared in the lower qubit state for $t=0$. The black solid line
corresponds to  the
time dependent 
probability for the Device 1 and red dashed line corresponds to the time
dependent 
probability for the Device 2. (a) $A_0=0.27$ meV/nm, (b) $A_0=25.0$ meV/nm and
(c)
$A_0=52.5$ meV/nm. 
}
\end{figure}

Equation~\ref{eq:et1} shows that, even if the driving is
in resonance with the frequency $\omega_{res}=(E_{q2} - E_{q1})/\hbar$, there
are
transitions to all the bounded states allowed by the dipole selection rules so,
unless somehow the matrix elements $Z_{kn}$ preclude this possibility, for
large enough time all the terms in the superposition
Equation~\ref{eq:superposition}
will have non-negligible contributions. This fact, inevitably, produces  a large
and undesirable leakage.

The points made above suggest that
a proper design of
the nano-structure would lead to negligible matrix elements $Z_{kn}$ for the
unwanted
transitions. This is the reason to choose structures that single out a couple
of eigenstates whose spatial extent is quite different from all the other
ones since, clearly, this is a economical way to reduce the transitions to 
eigenstates  that are not those of the qubit. The naive picture that says that
the basis states of a qubit can be chosen as two well separated states of a
physical system is really hard to be found, in particular when the
requirements of fast enough operations also must be accomplished
\cite{Ferron2012}. From a physical point of view, to obtain faster operation
times it is necessary to draw on stronger external drivings. By two well
separated states, it is meant that the energy difference between any other
state with the qubit states is  much larger than the energy difference
between the qubit states.

The matrix elements $Z_{kn}$ can be obtained for all the bounded states of
Devices 1 and 2 with very high precision so, the time evolution of the
electronic quantum state results from the  integration of Equation~\ref{eq:et1}.
The 
numerical 
integration was performed using high precision Runge-Kutta algorithms, taking
into account all the bounded states that each device possess.

Figure~\ref{fig:four} shows the time evolution of the occupation probability of
one of the qubit basis states, the initial condition is $\Psi(t=0)= |q\rangle$.
The Figure shows the time evolution for three different driving strengths and
two Devices, those whose eigenstates electronic densities are shown in
Figure~\ref{fig:three}. Clearly, as the driving increases its value, the time
evolution of the state becomes less and less harmonic. Anyway, for all the
driving strengths shown, a fast switching between states can be easily
achieved. The departure from a simple oscillatory behavior observed for larger
driving strengths, and that $|c_q(t)|^2<1$ for $t>0$, shows that the driving
is producing a superposition of many different eigenstates. On the other hand,
from the three panels of Figure~\ref{fig:four}, the dependence of the switching
time on the driving strength is manifest. 

As has been said above, that  $|c_q(t)|^2$ does not reach the unity for $t>0$,
is  manifestation of the probability leakage, {\em i.e.} the state does not
switch perfectly between the two qubit basis states. Anyway, for systems with a
finite number of states, there is a finite probability that the state of the
system  returns to the initial state for a large enough evolution time. A
potential well has indeed a finite number of eigenstates and, in many cases, the
numerical integration of Equations~\ref{eq:et1} imposes a further reduction of
the number of eigenstates effectively considered. For these reasons it is
useful to introduce time-averaged quantities to qualify the dynamical behavior
of the system.

\begin{figure}
\begin{center}
\includegraphics[scale=0.4]{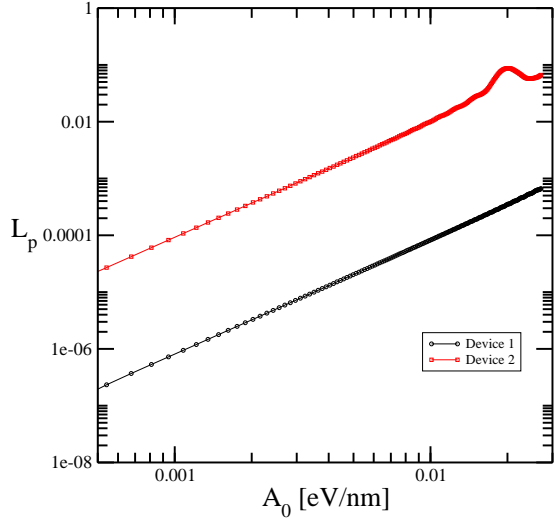}
\end{center}
\caption{\label{fig:five}(Color on-line) Time-averaged leakage $L_p$ for an rf
field pulse of strength $A_0$ at the resonant frequency 
$\omega_{res}=(E_{q2} - E_{q1})/\hbar$ as a function of the pulse strength for
the
Device 1
(black circle dots and line)
and the Device 2 (red squared dots and line).
}
\end{figure}

Together with the instantaneous leakage, $1 -(|c_{q1}|^2+c_{q2}|^2)$, it is
customary to introduce the time-averaged Leakage, $L_p$, which is defined as
\begin{equation}
 L_p = \frac{1}{T}\int_t^{t+T} \left(1
-(|c_{q1}(t^\prime)|^2+c_{q2}(t^\prime)|^2)
\right)\, dt^\prime ,
\end{equation}
where $T$ is a large enough time that, in principle, can be taken equal
to several
periods of the external driving. Figure~\ref{fig:five} shows the
behavior of the time-averaged Leakage for the two Devices previously defined as
a function of the strength of the external driving. As can be appreciated, the
time-averaged Leakage depends quadratically on the driving strength. On the 
other hand, 
Figure~\ref{fig:five} shows that, despite that Device 2 differs from Device 1
in just one single eigenstate localized in the innermost potential well, the
Leakage of both devices differ in two orders of magnitude. It is worth to
remark here that the spectrum of both devices are, essentially, the same (see
Figure~\ref{fig:two}).  

The good performance of Device 1 as a qubit can be further
emphasized, looking at the dynamical behavior of the quantum state, when the
system is forced with an off-resonance driving. 

\begin{figure}
\begin{center}
\includegraphics[scale=0.4]{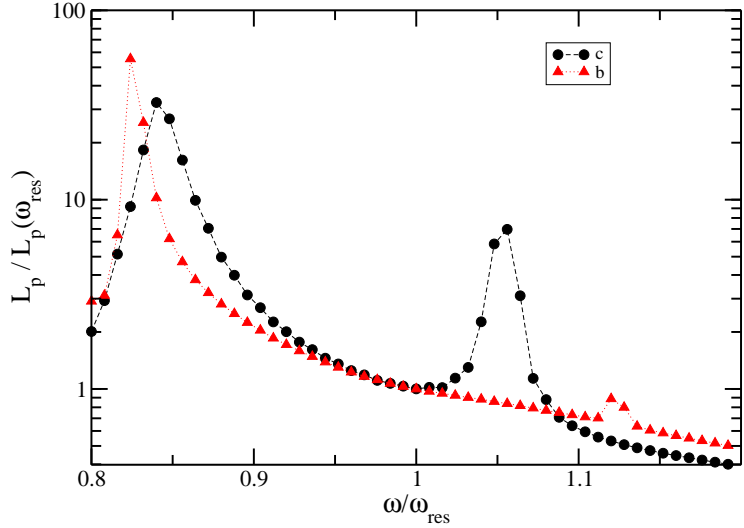}
\end{center}
\caption{\label{fig:six}(Color on-line) The Leakage experimented by the Device
1 when the external driving is off-resonance. Note that both axis scale are
normalized to the on-resonance values. The (red) triangle dots correspond to
the a driving strength of $A_0=25.0$ meV/nm and the (black) solid dots to
$A_0=52.5$ meV/nm, respectively. These driving strengths correspond to the cases
b) and c) of Figure~\ref{fig:four}
}
\end{figure}

Figure~\ref{fig:six} shows the behavior of the Leakage as a function of the
frequency of the external driving. It is assumed that the driving potential has
the same properties that the one in Equation~\ref{eq:drive}. The Figure shows
the data obtained for the two larger driving strengths showed in
Figure~\ref{fig:four} since these two are the  worse cases. As can be
appreciated, for the smaller driving strength the driving frequency can be
off-resonance up to $\pm 10\%$ without changing appreciably the Leakage. The
size of this tolerance interval is closely related to the ratio between the
eigen-energies differences with the driving strength.
For the larger driving, the tolerance
interval for frequencies smaller than $\omega_{res} $ is, at least, as larger as
the tolerance interval for the smaller driving. The peaks in both sets of data
signal that the transition probability to other states, which are not those of
the qubit, gets bigger, spoiling the switching between the qubit states. 

It is
worth to mention here that when the architecture of the quantum dot separates
two eigenstates, the residual leakage is produced by the small overlap between
the qubit states and all the other eigenstates. Actually, because the small
barrier that separates the potential wells, there is a non-negligible portion
of the qubit states that lies in the outermost well. A better design would
reduce or eliminate this portion enhancing the good behavior of the qubit. In
the next Section we present a different model potential that possess exactly all
the desirable properties that we have mentioned so far: separates in a potential
well the two lowest lying eigenstates, and reduces the overlap between the qubit
states and the other states to almost negligible values. Regrettably the model
potential, and the parameters that characterize it, does not correspond to a
nano-device already proposed. Anyway, since we have tested only a very limited
set of options, it is possible that the discussion of the ideas presented here
helps to find better designs for qubits based on multi-layered quantum dots.

\section{Other systems}
\label{othersys}

So far, we have not been  able to find an adequate set of parameters (radii,
materials and so on) to design a quantum dot model that separates  as qubit
states eigenstates with $n_r=0$.
The state with
$n_r=\ell=0$ has too many advantages to rule out too fast the search of a
quantum dot such that one of the two separated eigenstates be it. By separated
states we mean that two eigenstates are well localized in a given 
potential well, while all the other bounded eigenstates are localized in other
potential wells.  

\begin{figure}[floatfix]
\begin{center}
\includegraphics[scale=0.5]{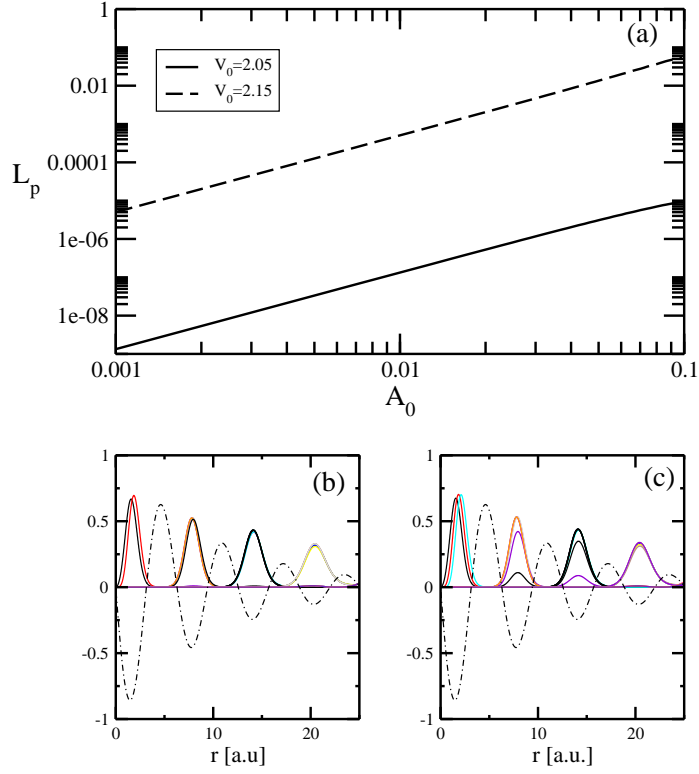}
\end{center}
\caption{\label{fig:seven}(Color on-line) a) The Leakage {\em vs} the driving
strength $A_0$, the value of $V_o$ is shown in the figure, while $\omega=1$ and
$\gamma=0.1$. The black solid line corresponds to the device shown in panel
b) and the dashed black line to the device shown in panel c). Note that the
devices differ in the number of eigenstates localized in the innermost well,
two and three states in panels b) and c), respectively. In panels b) and c) the
black dashed line represents the decaying sinusoidal potential, in units of
$A$. The color lines correspond to the electronic densities of several
eigenstates.
}
\end{figure}

Anyway, the separation of the lowest lying eigenvalue,$n_r=\ell=0$ , and an
excited state with $\ell=1$ can be achieved in the potential
\begin{equation}\label{eq:expot}
 V_c(r) = - V_0 e^{-\gamma r} \sin{(\omega r)},
\end{equation}
where $V_0$, $\gamma$ and $\omega$ are constants. The potential in 
Equation~\ref{eq:expot} has some common properties with step-like potentials
like the one in Equation~\ref{potrc} \cite{Ferron2012}. The parameters of the
potential and its shape are quite different from those commonly found in
nano-devices, so in this Section we use atomic units.

Figure~\ref{fig:seven} shows the Leakage suffered by the two ``devices''
depicted in panels b) and c) that has two and three states localized in the
innermost well, respectively. The parameter that drives the change from two to
three separated states is, in this case, $V_0$. The two lowest lying
eigenvalues have quantum numbers $(n_r=0,\ell=0)$ and $(n_r=0,\ell=1)$. We call
$V_0^c$ the critical value such that for $V_0<V_0^c$ there are only two well
localized states in the innermost well. For $V_0>V_0^c$ there are more than two
localized states in the innermost well, in our case $V_0^c\approx 2.09$

The eigenvalues and
eigenstates of the one-electron Hamiltonian with potential \ref{eq:expot} were
obtained approximately using the FEM-DVR method (finite-element method plus
discrete variable representation), see \cite{Lin2012,Tao2009,Rescigno2000}.

As can be seen in Figure~\ref{fig:seven} a) the Leakage of the
device with three separated states is larger than the leakage suffered by the
device with only two states separated. Again, the Leakage is a quadratic
function of the driving strength, $A_0$.

\begin{figure}[floatfix]
\begin{center}
\includegraphics[scale=0.4]{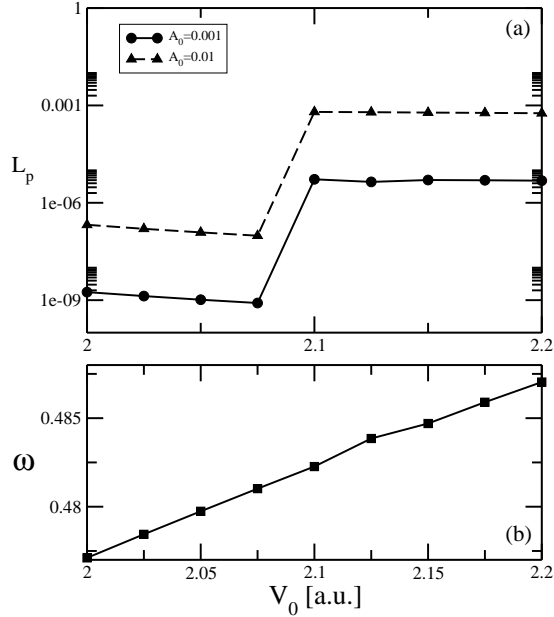}
\end{center}
\caption{\label{fig:eight}(a) Time-averaged leakage $L_p$ for an
rf field pulse of strength
$A_0=0.001$ (black solid dots and line) and $A_0=0.01$ (triangle dots and
dashed line), at the resonance frequency, as a function of $V_0$.
(b) The resonance frequency as a function of $V_0$.
}
\end{figure}

The influence of the number of separated states in the Leakage can be put
further in evidence analyzing its behavior when $V_0$ is swept from a 
smaller value than the critical to a  larger value than $V_0^c$. This
behavior is shown in Figure~\ref{fig:eight}. As can be appreciated from panel a)
there is a sudden change in the Leakage around the critical value, in fact the
Leakage changes three orders of magnitude when $V_0$ goes from smaller values
than the critical to larger ones, irrespective of the driving strength, at
least for the range of values analyzed. The jump can be rightly  attributed to
the presence of other states than those of the qubit since the others
quantities involved (in the Leakage) are continuous, for example,
Figure~\ref{fig:eight} b) shows the behavior of the resonance frequency as a
function of $V_0$.  

\section{Discussion and Conclusions}
\label{dissc}

The ability of multi- wells and barriers spherical potentials to separate a
subset of eigenstates in a potential well could be a tool to better designed
nano-devices. This capacity has not been yet systematized, there are not exact
results or theorems that allow a systematic search of model potentials with all
the desirable characteristics, despite it seems fairly general.

As has been said above, once two eigenstates are separated in a potential
well from all the other eigenstates, the residual leakage observed during the
switching between them is attributable to
the overlap between the qubit states and the other eigenstates. A higher
barrier between the potential wells could reduce the overlap but, so far, we do
not know of quantum dots built from three different semiconductor compounds.

The advantage of the ground state as one of the qubit basis state is obvious, it
can be pinpointed by cooling methods while other states require more
sophisticated means to force the electron to actually occupy one of them.

Our results contribute to show that there are, yet, a lot of improvements that
can be made to the design of qubits based on semiconductor quantum dots. The
huge amount of materials and geometries offer ample possibilities to tackle the
drawbacks that have marred the development of a reliable quantum dot based
qubit. Of course the decoherence induced by the spin-orbit interaction, not
considered in this work, stills remains as the heavier challenge. To minimize
the effects of spin-orbit interaction the qubit states should be located in a
potential well made of the material with the lowest spin-orbit interaction
strength possible.

\section*{Acknowledgements}
We would like to acknowledge  SECYT-UNC,  CONICET and MinCyT C\'ordoba
for partial financial support of this project. A.F. likes to acknowledge 
PICT-2011-0472 for partial financial support of this project. O.O. likes to
acknowledge the hospitality at the Instituto de Modelado e Innovaci\'on
Tecnol\'ogica
(CONICET-UNNE), where this project started.

\end{document}